%% file: surface_conference_jul15.tex
\documentclass[twocolumn,10pt]{asme2e}
\usepackage[T1]{fontenc}
\usepackage[latin9]{inputenc}
\usepackage{amsmath}
\usepackage{graphicx}
\usepackage{psfrag}
\usepackage{color}

\confshortname{ITP-09}
\conffullname{Interdisciplinary Transport Phenomena VI}
\confdate{4-9}
\confmonth{October}
\confyear{2009}
\confcity{Volterra} 
\confcountry{Italy}

\papernum{ITP-09-45}      

\title{A MIXED BASIS PERTURBATION APPROACH TO APPROXIMATE THE SPECTRUM OF LAPLACE OPERATOR}

\author{Matias Nordin\thanks{Address all correspondence to this author.}
    \affiliation{Applied Surface Chemistry \\
Department of Chemical and \\ Biological Engineering \\
Chalmers University of Technology \\
412 96 Gothenburg \\
Sweden \\
matias@chalmers.se
}
}

\author{Martin Nilsson-Jacobi
    \affiliation{Complex Systems Group \\
Department of Energy and Environment \\ 
Chalmers University of Technology \\
412 96 Gothenburg \\
Sweden \\
mjacobi@chalmers.se
}	
}

\author{Magnus Nyden
    \affiliation{Applied Surface Chemistry \\
Department of Chemical and \\ Biological Engineering \\
Chalmers University of Technology \\
412 96 Gothenburg \\
Sweden
}
}

\begin{document}
\maketitle    

\begin{abstract}
This paper presents a mixed basis approach for Laplace eigenvalue problems, which treats the boundary as a perturbation of the free Laplace operator. The method separates the boundary from the volume via a generic function that can be pre-calculated and thereby effectively reduces the complexity of the problem to a calculation over the surface. Several eigenvalues are retrieved simultaneously. The method is applied to several $2$ dimensional geometries with Neumann boundary conditions.
\end{abstract}

\begin{nomenclature}
\entry{$D$}{Diffusion constant.}
\entry{$\Delta$}{Laplace operator/Laplace matrix.}
\entry{$\lambda$}{Eigenvalue of $\Delta$}
\entry{$\Delta_F$}{Free Laplace operator.}
\entry{$S$}{Surface operator.}
\entry{$\Phi(x)$}{Fundamental solution of the Laplace equation.}
\entry{$\sigma_i(x)$}{Surface charge distribution.}
\entry{$\Phi_i(x)$}{Solution to the charge distribution $\sigma_i$.}
\entry{$\Psi_i$}{Eigenfunction to $\Delta_F$.}
\end{nomenclature}
\section*{INTRODUCTION} 
In physics and chemistry many situations with non-trivial boundary conditions give rise to a Laplace eigenvalue problem that is not analytically solvable. In such situations numerical methods, 
such as finite element methods (FEM)~\cite{bathe1995}, fast multi-pole methods (FMM)~\cite{greengard1987}, analytic element methods (AEM)~\cite{strack1999}, boundary element methods (BEM)~\cite{banerjee1994}, and boundary approximation methods (BAM)~\cite{li1987} (among many others) are used. In this paper an alternative method is presented that uses a perturbation expressed in a mixed basis which effectively reduces the complexity of the problem to a calculation over the surface. The mixed basis approach shares similarities with AEM,BEM and BAM, which also are formulated on the boundary, but has the advantage of not involving a fully populated matrix over the surface and results in an approximation of several eigenvalues of the Laplace spectrum. The method is applied on diffusion in 2 dimensions for various domains.

\section*{GENERAL CONSIDERATIONS}
We let $f$ be a real valued twice-differentiable function in an $n$-dimensional (Euclidean) domain. The diffusion equation is then defined as
\begin{equation}
	\partial _t v = D \Delta v . 
\end{equation}
The general approach of solving the diffusion equation  is to express the solution as a linear combination (a Schmidt decomposition) of terms that separate into a time-dependent and a spatially dependent part $v(t, \mathbf{x} ) = \sum _i \alpha _i f _i (\mathbf{x}) g_i (t)$. 
The spatial modes are defined by the (Laplace) eigenproblem
\begin{equation}
\Delta f_i = \lambda _i f_i
\label{eq:eig}
\end{equation}
where $\lambda_i$ is the separative constant. The collection of all constants $\lambda _i$ that solve this equation give rise to an eigenvalue spectrum of the operator $\Delta$. This spectrum depends on the boundary conditions used and on the shape of the boundary. 
In this paper we use Neumann boundary conditions
\begin{equation}
\nabla f _i (\partial \Omega)\cdot \hat{\mathbf{n}}=0
\label{eq:boundary}
\end{equation}
where $\partial \Omega$ denotes the boundary of the domain and $\hat{\mathbf{n}}$ its normal. With Neumann conditions the Laplace operator spectrum (also called the Neumann spectrum) consists of non-positive eigenvalues. In the absence of the boundary conditions we denote the free Laplace operator by $\Delta_F$ and recall that the eigenfunctions to $\Delta_F$ are harmonic functions.


\section*{SURFACE EXPANSION}

Our goal is to find an efficient computational tool to approximate part of the spectrum defined in Eq.~\ref{eq:eig}. A naive approach would be to treat the obstructing surface as a perturbation of the free Laplace operator, and use standard perturbation theory to derive the spectrum of the perturbed operator $\Delta$. Unfortunately this does not work because the surface obstruction cannot be expanded in any small parameter. In fact we can define a surface operator as

\begin{equation}
S = \Delta_{F} - \Delta ,
\label{eq:surf}
\end{equation}
where (as before) $\Delta _F$ denotes the free Laplacian. The norm of the operator $S$ is not small, and therefore standard perturbation theory does not work. 

In this paper we show that perturbation theory can still be useful, but that we need to find a more nontrivial set of basis functions. It is clear that, if the obstructions do not completely block the diffusion,  the large scale structure of the low frequency modes of $\Delta$ are dominated by eigenmodes of the free diffusion operator. More technically, we can say that for wavelengths significantly larger than the size of the obstruction, the obstructed diffusion operator recovers diffusive behavior, with an altered diffusion constant (i.e. shifted eigenvalues). We conclude that the eigenfunctions of the free Laplace operator should be part of the basis and that this part captures the large scale behavior of the eigenproblem. Furthermore we note that these eigenfunctions are known analytically and therefore there is no need to actually compute them.

To find the complementary basis that captures the small scale structure we use an adiabatic assumption. This is motivated by the fast relaxation of local variations during the  diffusion process. We treat the local variations close to the surface as fast degrees of freedom that can be treated using adiabatic elimination, i.e. we assume that the fast degrees of freedom equilibrate and are determined by the slow degrees of freedom. The local adiabatic equilibrium  of $\Delta$ is defined by Laplace equation $\Delta  \Phi = 0$  with a nontrivial boundary condition defined by the low frequency eigenmodes. In principle each low frequency eigenmode of the free Laplacian can be modified by adding an appropriate linear combination of solutions to Laplace equation so that the resulting function satisfies the von Neumann boundary condition. Here however we just add different solutions to the Laplace equation 
to the set of basis functions and use perturbation theory to calculate an approximate spectrum of the obstructed diffusion operator. The fundamental solution to the Laplace equation solves the problem 

\begin{equation}
\Delta \Phi ( \mathbf{x} ) = \delta ( \mathbf{x}_0 ) ,
\label{eq:fundSol}
\end{equation}
where $\delta$ is a Dirac delta function. In two dimensions the fundamental solution is $\log | \mathbf{x} - \mathbf{x}_0 |$ and in three dimensions it is $| \mathbf{x} - \mathbf{x}_0| ^{-1}$. A general solution of Laplace equation with some arbitrary (smooth) boundary condition can be expressed in terms of integrals over  fundamental solutions of different charge or dipole distributions over the surface, e.g.

\begin{equation}
	\Phi _i (x) = \int _S ds \; \sigma  _i (s) \log | \mathbf{x} - \mathbf{s} | ,
	\label{eq:surfaceIntegral1}
\end{equation}
for a charge distribution $\sigma  _i $ in two dimensions defined on a surface (one dimensional boundary).
%
The charge distribution is in general determined so that Eq.~\ref{eq:surfaceIntegral1} 
match the boundary condition (\ref{eq:boundary}). Here we generate basis functions from a set of smooth charge distributions. 

In this paper we are only considering two dimensional geometries, so the obstructions have a  one dimensional boundary that is straight forward to parametrize. The charge distributions are defined by low frequency Fourier modes, defined as  functions of a isometric parametrization on the surface of the obstructions. Furthermore, by placing the charges at a small distance outside the obstructing boundary and then mirroring a charge with opposite sign inside the boundary, we effectively create dipole charge distributions that are used to generate the basis functions. Parametrization of the obstructing surface in the three dimensional case it is more complicated but eigenmodes of the surface diffusion operator in Eq.~\ref{eq:surf} can be used.

Using the above construction we have defined a mixed basis consisting of solutions to the Laplace equation $\Phi _i$ and eigenfunctions to the free Laplace operator $\Psi _j$. To use perturbation theory the basis must be orthogonal. Since $\Delta _F$ is a self-adjoint operator, the free eigenfunctions can easily be constructed to be orthogonal to each other with respect to a scalar product over the entire volume. The basis from the Laplace equation must however be orthogonalized both with respect to themselves and to the free eigenfunctions. Using Eq.~\ref{eq:fundSol} we can express the scalar product between the free eigenfunction and the solutions to Laplace equation as a surface integral

\begin{eqnarray}
	\int _V dx \; \Psi _i (x) \Phi _j (x) & = & \frac{1}{\lambda ^F _i} \int _V dx \; \Psi _i (x) \Delta _F \Phi _j (x)\nonumber \\
	& =&  \frac{1}{\lambda ^F  _i} \int _S d y \; \Psi _i (y) \sigma ^d _j (y) .
\end{eqnarray}
Using Eq.~\ref{eq:surfaceIntegral1} the scalar product between two solutions to Laplace equation can be expressed as

\begin{equation}
	\int _V dx \Phi _i (x) \Phi _j (x)  =  \int _{S \times S} \!\!\! \!\!\!\!\!\! ds_1 ds_2 \; \sigma _i (s_1) \sigma _j (s_2) \Omega ( | s_1 - s_2| ) ,
\end{equation}
where
\begin{equation}
	\Omega ( |s_1 - s_2| ) = \int _V dx \; \log | s_1 - x | \log | s_2 - x| .
\end{equation}
The fact that $\Omega$ only depend on the distance between $s_1$ and $s_2$ can be realized through a rotation $x \rightarrow x'$ where $s_1$ and $s_2$ are placed on a generic, e.g. horizontal, axis at distance $|s-s'|$.  This rotation does not change the volume element $d x = d x'$. A key observation is that $\Omega$ is a generic function and can be pre-calculated and used in all the scalar products and also for different geometries. This means that the full orthogonalization of the mixed basis can be achieved using only surface integrals. 

Let $\Phi_i '$ denote linear combinations of $\Phi_i$ and $\Psi _j$, such that $\Phi _i '$ is orthonormal internally and to the free basis $\Psi _j$. Then perturbation matrix $A$, based on the orthogonalized mixed basis, has elements on the form $\int _V dx \Psi _i \Delta \Psi _j = \delta _{ij}$, $\int d x \; \Phi ' _i \Delta \Psi _j$, and $\int d x \; \Phi ' _i \Delta \Phi ' _j$. The elements  can be expressed as surface integrals using similar tricks as for the scalar products:

\begin{eqnarray}
	\int _V dx \; \Phi_i  (x) \Delta \Psi _j (x) & = & \int _S ds \; \sigma _i (s) \Psi _j (s) \\
	\int _V dx \; \Phi _i (x) \Delta \Phi _j (x) & = & \int _{S \times S} \!\!\! \!\!\!\!\!\!  ds_1 ds_2 \sigma _i (s_1) \sigma _j (s_2) \log | s_1 - s_2 | \label{eq:phiDphi} .
\end{eqnarray}

The eigenvalues of the matrix $A$ approximate the spectrum of $\Delta$ in Eq.~\ref{eq:eig} in the interval defined by the eigenvalues of the free eigenfunctions.  If we use a relatively small number of basis functions (typically less than $1000$), then the diagonalization of $A$ is very fast. In fact, the most computationally expensive step is the derivation of the elements in $A$, and especially the double surface integrals appearing in the scalar products involving two solutions to Laplace equation. To improve the computational costs we could possibly use the fact that the integral in  Eq.~\ref{eq:phiDphi} is identical to the total potential energy from two interacting charge distributions that do not interact internally. There exists several standard methods for efficient calculation of potential fields of this type, e.g. multi-pole methods\cite{greengard1987} and multi-grid methods\cite{brandt1977}. Using these methods it should be possible to reduce the cost of calculating the double surface integrals to $\mathcal{O} ( n_s \log ( n_s ) )$, where $n_s$ is the number of surface elements.

\section*{RESULTS AND DISCUSSION}
\begin{figure}[t]
\begin{center}
\input{circle}
\includegraphics[width=\columnwidth]{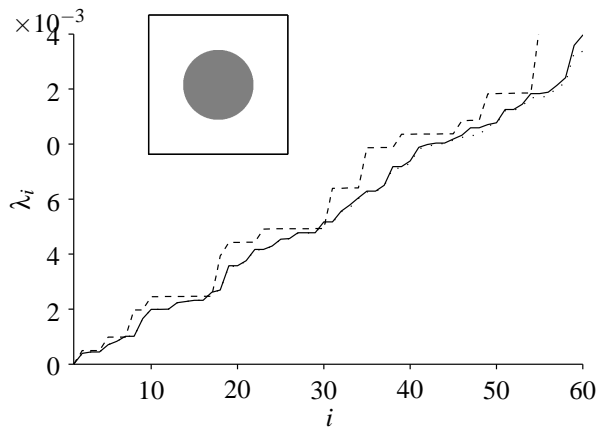} 
\end{center}
\caption{ORDINARY PERTURBATION (DASHED), SURFACE PERTURBATION WITH 100 FREE EIGENVECTORS AND 40 SURFACE EIGENVECTORS (SOLID) AND LAPLACE OPERATOR SPECTRUM (DOTTED). INSERTED PICTURE (ABOVE) SHOWS STRUCTURE}
\label{fig:circle} 
\end{figure}

\begin{figure}[t]
\begin{center}
\input{circuit}
\includegraphics[width=\columnwidth]{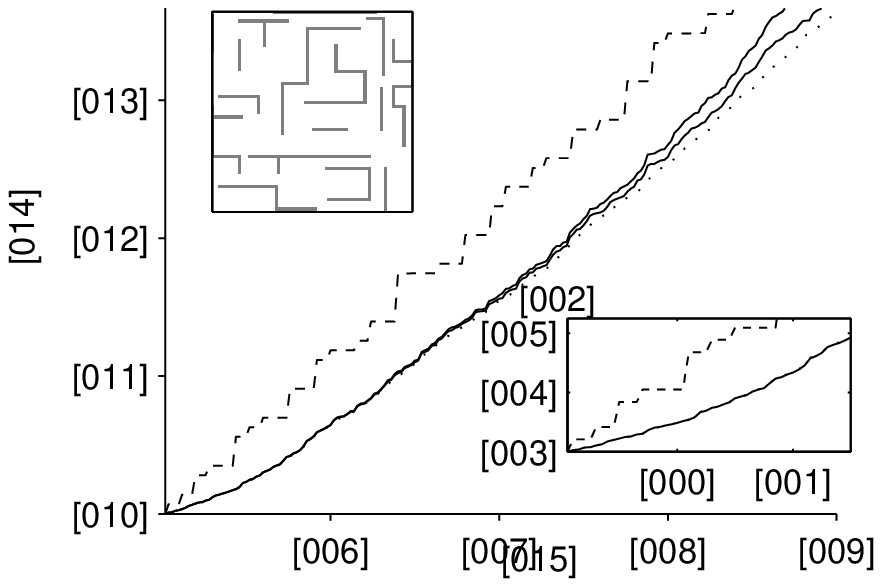} 
\end{center}
\caption{FREE SPECTRUM (DASHED), A SURFACE PERTURBATION WITH 200,300 FREE EIGENVECTORS AND 80 SURFACE EIGENVECTORS (SOLID) AND LAPLACE OPERATOR SPECTRUM (DOTTED). INSERTED PICTURE (ABOVE) SHOWS STRUCTURE AND (BELOW) ZOOM OF THE FIRST 50 EIGENVALUES.}
\label{fig:circuit} 
\end{figure}

\begin{figure}[b]
\begin{center}
\input{cox}
\includegraphics[width=\columnwidth]{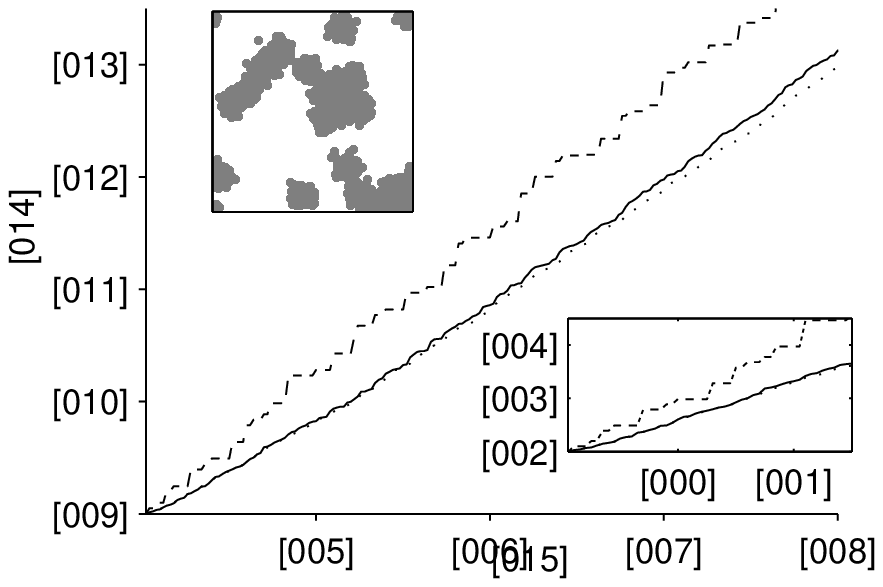} 
\end{center}
\caption{FREE SPECTRUM (DASHED), SURFACE PERTURBATION WITH 400 FREE EIGENVECTORS AND 250 SURFACE EIGENVECTORS (SOLID) AND LAPLACE OPERATOR SPECTRUM (DOTTED). INSERTED PICTURE (ABOVE) SHOWS STRUCTURE AND (BELOW) ZOOM OF THE FIRST 50 EIGENVALUES.}
\label{fig:cox} 
\end{figure}

\begin{figure}[b]
\begin{center}
\input{circle2}
\includegraphics[width=\columnwidth]{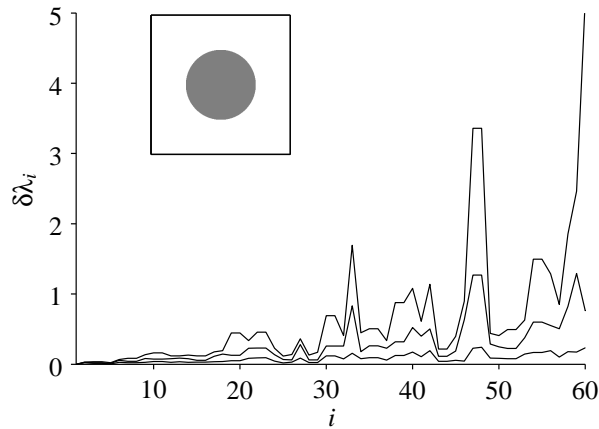} 
\end{center}
\caption{RELATIVE ERROR OF EIGENVALUES OF $A$ OF MIXED BASIS APPROACH, COMPARED TO REAL SPECTRUM OF $\Delta$. $A$ WAS CONSTRUCTED WITH 40 SURFACE EIGENVECTORS AND THE NUMBER OF FREE EIGENVECTORS WAS VARIED FROM 100 T0 200 T0 300. INSERTED PICTURE SHOWS STRUCTURE.}
\label{fig:circle2} 
\end{figure}

To test the surface expansion, the eigenvalue spectrum of the mixed matrix $A$ was calculated and compared with the original spectrum of the Laplace operator $\Delta$ for three different 2 dimensional geometries. The first example is a circle in a square domain with periodic boundary conditions. The domain consists of $160000$ elements and $A$ was created with the use of $100$ free eigenvectors and $40$ surface eigenvectors. Figure \ref{fig:circle} shows the resulting spectrum of the first $60$ eigenvalues in comparison with an ordinary perturbation calculation where $S$ is treated as a perturbation of $\Delta_F$, and the correct spectrum obtained by diagonalizing $\Delta$. The resultant spectrum of $A$ coincides well with the original spectrum of $\Delta$ and it is notable how poorly the naive perturbation captures the spectrum. 
The second example is a more complicated obstruction, a maze, which has a relatively large surface to volume ratio. The structure consists of $94249$ elements. 
Figure \ref{fig:circuit} shows the domain and the spectra of $\Delta_F$, $\Delta$ and $A$. The spectrum of $A$  was constructed from $80$ surface eigenvectors and between $100$ and $200$ free eigenvectors. The figure includes much useful information. The first eigenvalues of $A$ coincides well with the eigenvalues of $\Delta$, which implies that the basis of $A$ yields a good estimate of the low frequency part of the spectrum. Further up in the spectrum the eigenvalues are not accurately estimated, but with the use of more free eigenvectors the basis of $A$ is better capturing the spectrum of $\Delta$. This observation indicate that the communication between free eigenvectors, via the surface, is locally confined for the low frequency part of the spectrum. The internal communication of eigenvectors can be measured by investigating the eigenvectors of $A$ and proposes an error estimate by measuring the weight of the associated eigenvectors of $A$, where few elements of an eigenvector to $A$ implies a good estimate 
 of the eigenvalue. This has however not been studied in detail.  
Finally we present a randomly generated domain of $40000$ elements. In this domain a large number of surface eigenvectors was needed to achieve good results, probably due to the large number of sharp corners present. Figure \ref{fig:cox} shows the domain and the spectrum of $\Delta_F$, $\Delta$ and $A$, where $250$ surface eigenvectors and $400$ free eigenvectors where used. A resonant influence was found in the situations where the wave number of the eigenvectors coincided with the geometry studied. For such eigenvectors, the corresponding eigenvalues of $A$ converge slower to the actual eigenvalues with increasing number of free eigenvectors. This can be seen in figure \ref{fig:circle2} where the relative error of the eigenvalues of $A$ are plotted against their index for the circular obstruction domain. The peaks correspond to wave numbers that coincide with the radius of the circle.

\section*{SUMMARY AND CONCLUSION}
We have constructed a mixed basis that can be used in perturbation calculations of the spectrum of the Laplace operator in complicated geometries. This reduces the problem to a generic function over the volume, which can be pre-calculated, and surface integrals. Relatively few vectors are needed to span the perturbation matrix which yields for a computationally interesting approach of solving problems involving the spectrum of the Laplace operator, e.g. diffusion. Existing techniques such as FEM, AEM, BEM and BAM gives good estimates of the Laplace spectrum but generally involve many elements. AEM, BEM and BAM are closely related to our method, as the calculations are also formulated on the surface, but for AEM and BEM; require a full diagonalization over the surface, and for BAM; an iterative scheme, which is needed for each eigenvalue. The mixed basis approach gives the opportunity of reducing the size of the resultant matrix to derive an approximation to a part of the spectrum. We have demonstrated that the resulting spectrum 
 is a good estimate and that relatively few elements are needed for good results. We have not yet compared the performance of the new method to existing techniques.  A possible application of our method could be to initialize the solution scheme for an iterative eigenvalue/eigenfunction solver, for example the method by Li et al~\cite{li2006} that relies on an initial guess on the eigenvalues.

\begin{acknowledgment}
Nyd\'en would like to thank the Vinova financed VINN Excellence Centre SuMo Biomaterials (Supermolecular Biomaterials- Structure dynamics and properties) and Nordin would like to thank the Swedish Science Council (VR project no. 2008-3895) for funding.
\end{acknowledgment}

%
% ...

\bibliographystyle{ieeetr}
\bibliography{references}

\end{document}

%% file: circle.tex
%
%
\providecommand\matlabtextA{\color[rgb]{0.000,0.000,0.000}\fontsize{10}{10}\selectfont}%
\psfrag{[011]}[bc][bl]{\matlabtextA $\lambda_i$}%
\psfrag{[012]}[tc][bl]{\matlabtextA $i$}%
%
%
%
\psfrag{[000]}[ct][ct]{\matlabtextA $10$}%
\psfrag{[001]}[ct][ct]{\matlabtextA $20$}%
\psfrag{[002]}[ct][ct]{\matlabtextA $30$}%
\psfrag{[003]}[ct][ct]{\matlabtextA $40$}%
\psfrag{[004]}[ct][ct]{\matlabtextA $50$}%
\psfrag{[005]}[ct][ct]{\matlabtextA $60$}%
%
%
%
\psfrag{[006]}[br][br]{\matlabtextA $\times10^{-3}$}%
%
%
%
\psfrag{[007]}[rc][rc]{\matlabtextA $0$}%
\psfrag{[008]}[rc][rc]{\matlabtextA $2$}%
\psfrag{[009]}[rc][rc]{\matlabtextA $4$}%
\psfrag{[010]}[rc][rc]{\matlabtextA $6$}%
%

%% file: circuit.tex
%
%
\providecommand\matlabtextA{\color[rgb]{0.000,0.000,0.000}\fontsize{10}{10}\selectfont}%
\psfrag{[014]}[bc][bl]{\matlabtextA $\lambda_i$}%
\psfrag{[015]}[tc][bl]{\matlabtextA $i$}%
%
%
%
\psfrag{[000]}[ct][ct]{\matlabtextA $20$}%
\psfrag{[001]}[ct][ct]{\matlabtextA $40$}%
\psfrag{[006]}[ct][ct]{\matlabtextA $50$}%
\psfrag{[007]}[ct][ct]{\matlabtextA $100$}%
\psfrag{[008]}[ct][ct]{\matlabtextA $150$}%
\psfrag{[009]}[ct][ct]{\matlabtextA $200$}%
%
%
%
\psfrag{[002]}[br][br]{\matlabtextA $\times10^{-3}$}%
%
%
%
\psfrag{[003]}[rc][rc]{\matlabtextA $0$}%
\psfrag{[004]}[rc][rc]{\matlabtextA $2$}%
\psfrag{[005]}[rc][rc]{\matlabtextA $4$}%
\psfrag{[010]}[rc][rc]{\matlabtextA $0$}%
\psfrag{[011]}[rc][rc]{\matlabtextA $0.006$}%
\psfrag{[012]}[rc][rc]{\matlabtextA $0.012$}%
\psfrag{[013]}[rc][rc]{\matlabtextA $0.018$}%
%

%% file: cox.tex
%
%
\providecommand\matlabtextA{\color[rgb]{0.000,0.000,0.000}\fontsize{10}{10}\selectfont}%
\psfrag{[014]}[bc][bl]{\matlabtextA $\lambda_i$}%
\psfrag{[015]}[tc][bl]{\matlabtextA $i$}%
%
%
%
\psfrag{[000]}[ct][ct]{\matlabtextA $20$}%
\psfrag{[001]}[ct][ct]{\matlabtextA $40$}%
\psfrag{[005]}[ct][ct]{\matlabtextA $50$}%
\psfrag{[006]}[ct][ct]{\matlabtextA $100$}%
\psfrag{[007]}[ct][ct]{\matlabtextA $150$}%
\psfrag{[008]}[ct][ct]{\matlabtextA $200$}%
%
%
%
\psfrag{[002]}[rc][rc]{\matlabtextA $0$}%
\psfrag{[003]}[rc][rc]{\matlabtextA $0.01$}%
\psfrag{[004]}[rc][rc]{\matlabtextA $0.02$}%
\psfrag{[009]}[rc][rc]{\matlabtextA $0$}%
\psfrag{[010]}[rc][rc]{\matlabtextA $0.02$}%
\psfrag{[011]}[rc][rc]{\matlabtextA $0.04$}%
\psfrag{[012]}[rc][rc]{\matlabtextA $0.06$}%
\psfrag{[013]}[rc][rc]{\matlabtextA $0.08$}%
%

%% file: circle2.tex
%
%
\providecommand\matlabtextA{\color[rgb]{0.000,0.000,0.000}\fontsize{10}{10}\selectfont}%
\psfrag{[012]}[bc][bl]{\matlabtextA $\delta \lambda_i$}%
\psfrag{[013]}[tc][bl]{\matlabtextA $i$}%
%
%
%
\psfrag{[000]}[ct][ct]{\matlabtextA $10$}%
\psfrag{[001]}[ct][ct]{\matlabtextA $20$}%
\psfrag{[002]}[ct][ct]{\matlabtextA $30$}%
\psfrag{[003]}[ct][ct]{\matlabtextA $40$}%
\psfrag{[004]}[ct][ct]{\matlabtextA $50$}%
\psfrag{[005]}[ct][ct]{\matlabtextA $60$}%
%
%
%
\psfrag{[006]}[rc][rc]{\matlabtextA $0$}%
\psfrag{[007]}[rc][rc]{\matlabtextA $1$}%
\psfrag{[008]}[rc][rc]{\matlabtextA $2$}%
\psfrag{[009]}[rc][rc]{\matlabtextA $3$}%
\psfrag{[010]}[rc][rc]{\matlabtextA $4$}%
\psfrag{[011]}[rc][rc]{\matlabtextA $5$}%
%